\documentclass[12pt]{iopart}

\usepackage{color}
\usepackage{epsf}
\usepackage{graphicx}
\usepackage{iopams}  
\begin{document}

%\begin{frontmatter}

%\pagenumbering{arabic}

\title[Electronic Structure of FeAl Alloy]{Electronic Structure of FeAl Alloy Studied by Resonant Photoemission Spectroscopy and {\textit{Ab Initio}} Calculations} 

\author{Debashis Mondal$^a$, Soma Banik$^{b,*}$, C. Kamal$^b$, Mangla Nand$^c$, S. N. Jha$^c$, D. M. Phase$^d$, A. K. Sinha$^{a,b}$, Aparna Chakrabarti$^{a,b}$, A. Banerjee$^d$ and Tapas Ganguli$^{a,b}$}
\address{$^a$Homi Bhabha National Institute, Raja Ramanna Centre for Advanced Technology, Indore, 452013, India.}
\address{$^b$Indus Synchrotrons Utilization Division, Raja Ramanna Centre for Advanced Technology, Indore, 452013, India.}
\address{$^c$Atomic and Molecular Physics Division, Bhabha Atomic Research Center, Mumbai, 400085, India.}
\address{$^d$UGC-DAE Consortium for Scientific Research, University Campus, Khandwa Road, Indore, 452001,India.}
\ead{$^*$soma@rrcat.gov.in}

\begin{abstract} 
Resonant photoemission spectroscopy has been used to  investigate the character of Fe $3d$ states in FeAl alloy. Fe $3d$ states have two different character,  first is of itinerant  nature located very close to the Fermi level, and second, is of less itinerant (relatively localized character),   located beyond $\sim$2~eV below the Fermi level.  These distinct states are clearly distinguishable in the resonant photoemission data. Comparison between the results obtained from experiments and first principle based electronic structure calculation show that the origin of the itinerant character of the Fe $3d$ states is due to the ordered $B2$ structure, whereas the relatively less itinerant (localized) Fe $3d$ states are from the disorders  present in the sample. The exchange splitting of the Fe $3s$ core level peak confirms the presence of local moment in this system. It is found that the itinerant electrons arise due to the hybridization between Fe $3d$ and Al $3s-3p$ states. Presence of hybridization is observed as a shift in the Al $2p$ core-level spectra as well as in the X-ray near edge absorption spectra towards lower binding energy.  Our photoemission results are thus explained by the co-existence of ordered and disordered phases in the system.   
\end{abstract}

\maketitle

\section{Introduction}

FeAl intermetallic alloys show very attractive physical properties like high mechanical strength and excellent corrosion resistance {\it etc}\cite{Kowalski, Tortorelli}. These properties make FeAl alloys a potential candidate for structural and coating applications and prove it to be a promising substitute for stainless steel\cite{Stoloff, Liu, Zeng}. Stoichiometric FeAl shows either an ordered $B2$ phase or a disordered $A2$ phase.  $Fe_{1-x}Al_x$ for $x~<$~0.5 shows other ordered magnetic phases like $D03$ and $B32$ \cite{Phasedig, Becker}. The structural characteristics of these alloys have been related to the magnetic  properties and different phases like ferromagnetic (FM), paramagnetic (PM), spin-glass (SG) and re-entrant spin glass  are reported \cite{Phasedig, Becker, Parthasarathi, Domke, Bogner}. X-ray photoemission (XPS) valence band of FeAl multilayer sample showed appreciable changes in the Fe $3d$ and the Al $3s$ density of states as a function of annealing which are attributed to the strong $sp-d$ hybridization at the Fermi level ($E_F$) \cite {Brajpuriya09}. Oku {\it et al.}\cite{oku06} reported a magnetic moment of 1.7 $\mu_B$  in the PM phase from the analysis of Fe $3s$ XPS core level spectra. They attributed the presence of moments in this system to the magnetic ordering in the sub-surface layers. In another communication, Oku {\it et al.}\cite{Oku06_JESRP} reported that the X-ray photoemission spectra of FeAl in PM state could not be explained by electronic structure of the ground state alone. 

Electronic-structure  calculations  based  on the local density approximation (LDA) using density functional theory (DFT) on ideal B2 phase of FeAl showed a FM ground state with $\sim$0.7~$\mu_B$ \cite {Bogner, Williams, Chacham, Sundararajan, Kulikov}.  The calculations based on the disordered local moment theory \cite{Bose} is also unable to explain the observed PM state. More recently, a NM ground state has been obtained for ideal B2-FeAl by empirically choosing the parameter U=~5~eV within the LDA + U method \cite{Mohn, Petukhov}. Using Korringa, Kohn, and Rostoker method, it has been shown that the partial long range order due to the thermal anti-sites could be responsible for the paramagnetic state in this system \cite{Smirnov}. So there is a long standing question to explain the magnetic state observed in the experiment. Neither ferromagnetic $LDA$ nor non-magnetic LDA + U  results  explain the existing experimental data. Hence, it is utmost necessary to determine the electronic density of Fe states in FeAl both experimentally and theoretically.

 Although several calculations have been reported where the magnetic properties of FeAl have been studied, there are no reports where the calculated valence band DOS has been compared with the experimentally obtained total and partial density of states(DOS). Therefore in this work, we have carried out photoemission and resonant photoemission (RPES) measurements on FeAl and compared the results with theoretical calculations for different configurations like ordered and disordered FeAl structures. We have also carried out magnetic measurements on the same piece of sample and compared the results with the calculated magnetic moments and $3s$ exchange splitting moment. The important result of this work is that, we could be able to clearly distinguish between the itinerant and relatively localized character of the Fe $3d$ states which arise from the ordered and the disordered parts of the system respectively.

\section{Experimental and Computational Methods}

Polycrystalline ingot of FeAl was prepared by arc melting of 99.99\% pure constituent elements in an Ar atmosphere. The ingot was vacuum annealed in 7$\times$10$^{-7}$~mbar at 1100 $^\circ$C for 4 days, followed by quenching in ice water. Energy Dispersive Analysis of X-ray (EDAX) measurement has been carried out on the circular disc cut from the ingot.  The actual bulk composition determined from EDAX is Fe$_{0.49}$Al$_{0.51}$, which is close to the intended composition. Small pieces were cut from the ingot and then ground into fine powder for X-ray diffraction (XRD) measurements. XRD pattern from the powdered sample was recorded with a MAR 345 image plate detector at the angle-dispersive x-ray diffraction beamline (BL-12, Indus-2 synchrotron radiation source), using 19.3 keV x-ray photons in the forward scattering geometry. The wavelength and the sample to detector distance was determined using XRD pattern of LaB$_6$. FIT2D software was used to generate the XRD pattern from the diffraction rings obtained from the image plate data.  X-ray near edge spectroscopy (XANES) measurements were performed at the EXAFS beamline (BL-9, Indus-2 synchrotron radiation source).  XANES measurement was performed in fluoresence mode at Fe K-edge in the energy range from 7034 to 7193 eV. Magnetization and AC-suceptibility have been measured using a Quantum Design Superconducting Quantum Interference Device (SQUID) magnetometer. Magnetization measurements as a function of field at two different temperatures (2~K and 300~K) and the zero field cooled (ZFC) and field cooled (FC) magnetization at three different magnetic fields (0.01 T, 0.1 T and 1 T) have been carried out. The valence band (VB) normal and resonant photoemission measurements were performed at the Angle Resolved Photoelectron Spectroscopy beamline in Indus-1. The base pressure for the measurements was $\sim3~ \times 10^{-10}$ mbar and the analyser used was Phoibos 150. The samples were cleaned by Ar ion sputtering and the absence of Carbon $1s$ peak at ~ 287 eV and Oxygen $1s$ peak at 531~eV was ensured before the measurements. The VB photoemission spectra were recorded by varying photon excitation energies from 46~eV to 79~eV across the Fe $3p-3d$ and Al $2p-3s$ transition energies with the energy resolution of 135~meV. The Fermi edge of the sample was aligned to Fermi edge of Au for the same monochromator settings. XPS measurements were made in an Omicron system using an Al $k_\alpha$ source. 

To understand the experimental results on FeAl, we have performed DFT \cite{Hohenberg65} based spin-polarized electronic structure calculations by using Vienna Ab-initio Simulation Package (VASP)\cite{Kresse, Kresse2} within the framework of the projector augmented wave (PAW) method.  For exchange-correlation functional, we have used the generalized gradient approximation (GGA) given by Perdew, Burke, and Ernzerhof (PBE)\cite{Perdew}.  The energy cut-off of plane waves (basis set) is 400 eV. For Brillouin zone integration, we use Monkhorst-Pack scheme with k-meshes of 22$\times$22$\times$22. To probe the effect of chemical disorder on the properties of FeAl, we have also carried out the calculations for a (3$\times$3$\times$3) super cell (containing 27 Fe and 27 Al atoms) in which Fe and Al atoms are randomly arranged. For the super cell calculations, we have used the k-mesh size of 7$\times$7$\times$7 and the energy cut-off of 300~eV.  For these cases, we have performed the calculations in both magnetic and non-magnetic configurations. The convergence criteria for energy in SCF cycles is chosen to be 10$^{-6}$~eV. The geometric structures are optimized by minimizing the forces on individual atoms with the criterion that the total force on each atom is below 10$^{-2}$~eV/$\AA$. 

\section{Results and Discussions}

%\subsection{X-ray diffraction (XRD)} 
  
Room temperature XRD patterns recorded at 19.3~keV X-ray energy are shown in figure 1(a).  All the peaks could be indexed to a cubic structure using Le-Bail fitting, with a lattice constant of $a$=~2.917~$\AA$, which is in good agreement with the lattice parameter reported in literature \cite{Rajan, Ghosh, Apinaniz, Apinaniz1}. The presence of (100) peak in the pattern clearly indicates the presence of an ordered $B2$ phase in the sample.  However the intensity ratios of the various peaks in the diffraction pattern do not confirm the presence of the ordered $B2$ phase alone, thereby indicating presence of disordered phase in the sample. Further, a careful observation of the XRD pattern shows some small peaks marked by dots in figure 1(a). Such peaks have also been observed in References \cite{Rajan, Ghosh, Apinaniz1} and are ascribed to certain disordered phases, but a plausible explanation of their origin has not been ascertained.

%\begin{figure*}
%\center
%\epsfxsize=65mm
%\epsfxsize=100mm
%\epsffile{Fig1.eps}
%\vskip -0.3cm
%\caption {Color online: (a) X-ray diffraction pattern of FeAl at 300 K recorded with synchrotron source at 19.3~keV energy. The experimental data are denoted by open red circles, while the black solid line through the circles represents the calculated pattern. The lower dotted line represents the difference curve between experimental and calculated patterns. The dots in (a) represent the small impurity phases present in the sample. Magnetization as a function of applied magnetic field at 300~K and 2~K is shown in (b). Zoomed region of M(H) curve as in (b) near H=~0 is shown in inset for the hysteresis loop observed at 2~K. AC-susceptibility as a function of temperature is shown in (c) with the Blocking temperature and SG freezing temperature marked as $T_B$ and $T_F$ respectively.}
%\label{fig1}
%\vskip -0.2cm
%\end{figure*}

%\subsection{Magnetic susceptibility}
The magnetization as a function of field at temperature 2~K and 300~K are shown in figure 1(b). Both the M(H) curves do not saturate even at the maximum applied field of 7~Tesla. Similar behavior in M(H) curve has also been reported earlier in references \cite{Rajan, Lue}. M(H) curve at 300~K displays normal PM behavior, where as at 2~K (figure 1(b)), it shows a narrow hysteresis loop with a small coercivity $\sim$210~mT (as shown in figure 1(d)). This kind of behavior is reported in case of Au-Fe system \cite{Roy1, Roy2, Bitoh} and other FeAl alloy systems like FeAl$_2$, Fe$_2$Al$_5$ {\it etc} \cite{Jaglicic}. In figure 1(c) the susceptibility measurement is shown as a function of temperature. Two prominent transitions have been observed at 56~K and 24~K. The broad feature at 56~K can be attributed to the magnetic blocking effect induced by the competition between thermal energy and magnetic anisotropy energy due to the presence of magnetic clusters (introduced by the disorders) in the system (marked as T$_B$ in figure 1(c))\cite {Huang}. The transition at 24~K is relatively sharp which could be related to the SG freezing temperature (marked as T$_F$ in figure 1(c)) or some other magnetic impurity phase. This feature at 24 K is seen clearly in the M-T curve recorded at different magnetic field (figure 2(a)). Similar appearance of blocking as well as the SG transitions has been observed in the Ni ferrite nano-particles in SiO$_2$ matrix \cite{Nadeem}. Drawing an analogy with the mentioned references , we conclude that our sample also contains disordered phase  which may come from sample preparation condition \cite{Bogner, Rajan, PKM} (this is confirmed from XRD measurements shown in figure 1(a)). The presence of disorder where Al is replaced by Fe and vice versa is also expected to form small Fe clusters.

%\begin{figure}
%\center
%\epsfxsize=80mm
%\epsfxsize=90mm
%\epsffile{Fig2.eps}
%\vskip -0.3cm
%\caption {Color online:(a)Field Cooled (FC) and Zero Field Cooled (ZFC) magnetization measurement of FeAl under applied magnetic field of 0.01~T, 0.1~T and 1~T. (b) Temperature dependence of 1/$(\chi-\chi_0)$ for FeAl determined from the FC magnetization curves of 1~T and 0.1~T as shown in (a) (details are given in the text).}
%\label{fig2}
%\vskip -0.2cm
%\end{figure}

%\subsection{Magnetization}
A further confirmation of the presence of Fe clusters in the disordered FeAl system is obtained from the M-T curve recorded in the ZFC-FC protocol at three  different applied magnetic fields such as 0.01~T, 0.1~T and 1~T (shown in figure 2(a)). The ZFC and FC curve at 0.01~T bifurcates below 280~K, and this bifurcation temperature reduces significantly with increasing field. This clearly indicates the presence of magnetic in-homogeneity like the random presence of clusters. Similar bifurcation in ZFC and FC curves have been observed in another intermetallic system $Co_{0.2}Zn_{0.8}Fe_{1.95}Ho_{0.05}O_4$, where clustering has been reported \cite{Bhowmik}. The blocking effect ($T_B$=~56~K) get suppressed in these applied magnetic fields but the magnetic transition ($T_F$=~24~K) is quite prominent. The probable reason  for this kind of complex magnetism could be due to presence of magnetic clusters  and/or Fe chains formed easily in the finite disordered sample.  

The temperature dependent susceptibility and inverse susceptibility can be derived from the FC magnetization curve for H= 1~T and 0.1~T where the ZFC-FC branching (shown in figure 2(a)) in the paramagnetic regime is no more observable. The analysis of the susceptibility in the high temperature paramagnetic regime was performed using the Curie-Weiss law:
\begin{eqnarray}
	\chi~=~\chi_0~+~\frac{C}{T-\theta}
\end{eqnarray}
  where $\chi_0$ is the temperature-independent part of the susceptibility, $C$ is the Curie-Weiss constant and $\theta$ is the Curie-Weiss temperature. The constant $C$ gives information about the magnitude of the Fe moments. Above the ferromagnetic transition the data follow the Curie-Weiss law. The variation of ($\chi-\chi_0$)$^{-1}$ with respect to temperature is shown in figure 2(b). The fit of the Curie-Weiss law for susceptibility (solid line in figure 2(b)) yields the parameter values $\chi_0$=~8.23~$\times$~$10^{-4}$~emu/mol, $C$=~0.292 emu K/mol, and $\theta$=~-53~K. The effective moment is given as $\mu_{eff}$=~p$_{eff}$~$\mu_B$ per Fe ion \cite{Jaglicic} where $p_{eff}$=~$\sqrt{8C}$. Hence, the effective moment on the Fe atoms in this system is 1.5~$\mu_B$. 
$\chi_0$ has the contibution from the Larmor diamagnetic susceptibilities, the Pauli paramagnetic spin susceptibility and the Landau orbital diamagnetic susceptibility. In metals the diamagnetic response is negligible since the metals can be viewed as closed-shell ion cores and a free electron gas formed from the valence electrons. The magnetic behavior is then dominated by the Pauli paramagnetism of the conduction electrons. Pauli paramagnetic susceptibility is given as: $\chi^{Pauli}$=~10$^{-6}$$\frac{2.59}{r_s/a_0}$ where $r_s$ is the Wigner-Seitz radius and $a_0$ is the atomic Bohr radius \cite{Aschroft}. For metallic Fe, the value of $r_s/a_0$=~2.12 \cite{Aschroft} in the equation yields the $\chi^{Pauli}$=~1.22~$\times$10$^{-6}$ $\mu_B$. The value of $\chi_0$ is about $\approx$~675 times larger than $\chi^{Pauli}$ which implies that there should be small magnetic ordering present in the signal of the paramagnetic phase. This further confirms the presence of magnetic clusters in the system which are responsible for the splitting of the ZFC and FC magnetization curves.

%\subsection{X-ray photoelectron spectroscopy (XPS)} 
The XPS core levels of Fe $2p$, Fe $3p$, Al $2s$ and Al $2p$ of FeAl are compared with those of metallic Fe and Al which are shown in figures 3(a), (b), (c) and (d) respectively. Inelastic background has been subtracted from the core level spectrum using standard Tougaard method \cite{Tougaard}. The satellite contribution of Mg k$_{\alpha}$ source has also been subtracted from the core level spectrum taking into account that the satellite produces a replica spectrum as the main line, but shifted in energy and reduced in intensity \cite{Biswas}. The surface composition has been determined from the ratio of the area under the core level peaks Fe $2p$ and Al $2p$ normalized with the respective photoionization cross-section \cite{Yeh}. The surface composition is found to be Fe$_{0.6}$Al$_{0.4}$ which is slightly different from the bulk composition of Fe$_{0.49}$Al$_{0.51}$ as determined from EDAX. The difference in the composition obtained from EDAX and XPS measurements could be related to the errors in the estimation which are about 2$\%$ for EDAX and 10$\%$ in case of XPS. The Fe $2p$ core level in figure 3(a), shows a spin-orbit splitting of 13~eV with the Fe $2p_{1/2}$ and $2p_{3/2}$ peaks appearing at -719.4~eV and -706.4~eV, respectively. The Fe $3p$ core level in figure 3(b) for both Fe and FeAl appears at -52.25~eV. For Al metal the Al $2p$ and Al $2s$ peaks in figure 3(c) and (d) appear at -72.3~eV and -117.5~eV, respectively. Interestingly, the Al $2s$ and Al $2p$ peaks of FeAl show a shift of $\sim$ 0.4~eV towards lower binding energy (BE) than core level peaks of Al metal (in figure 3(c) and (d)). The BE of the core levels can show shifts depending on the chemical environment. The possibility of shift in BE peak can be assigned due to different phenomena like charge transfer \cite{Gelius74, Cole97}, heat of mixing \cite {Martensson81,Steiner81}, segregation energy \cite{Rosengren81}, and cohesive energy \cite{Johnson80}. The electro-negativity of Fe and Al are 1.83 and 1.61 respectively \cite{periodictable}. So Fe has an affinity to to attract the electron cloud towards itself. In that case Al core level peak should shift to higher BE but we observe the contrary. Hence, we predict that there is a charge redistribution in the system in such a way that the Fe electrons are being shared by the Al atoms in this system causing the shift of the Al core level to lower BE. This phenomenon will largely depend on the kind of hybridization present in the system \cite{Abrikosov01} and will be more evident in the valence band spectra.

%\begin{figure}
%\center
%\epsfxsize=90mm
%\epsfxsize=65mm
%\epsffile{Fig3.eps}
%\vskip -0.3cm
%\caption {Comparison of the core-level spectra of FeAl with Fe metal and Al metal showing (a) Fe $2p$, (b) Fe $3p$, (c) Al $2s$ and (d) Al $2p$ core levels. Dotted lines show the positions of the core level peaks.}
%\label{fig3}
%\vskip -0.2cm
%\end{figure}
 
%\subsection{Resonant photoemission spectroscopy} 
To study the nature of the Fe $3d$ states in FeAl we have recorded the VB spectra across the Fe $3p$ and Al $2p$ threshold energies and these spectra are shown in figure 4(a). The VB of FeAl in figure 4(a) shows three prominent features at -0.80, -3 and -6~eV marked as $A$, $B$ and $C$ respectively. Across the Fe $3p-3d$ resonance the VB spectra show clear changes in the features where antiresonance has been observed at 52~eV excitation energy (black spectra in figure 4(a)). Al $2p-3s$ transition is observed at 73~eV excitation energy (black spectra in figure 4(a)). At both the resonances we have observed substantial changes in the valence band features $A$, $B$ and $C$ which indicates that the Fe $3d$ states are hybridized with the Al $3s-3p$ states. 
 
%\begin{figure}
%\center
%\epsfxsize=100mm
%\epsfxsize=65mm
%\epsffile{Fig4.eps}
%\vskip -0.3cm
%\caption {VB spectra of FeAl recorded across the Fe $3p-3d$ transition and the Al $2p-3s$ transition are shown in (a) with the prominent features marked as A, B, C. The constant initial state spectra of the VB features at fixed binding energies are shown as a function of photon energy in (b). The solid line shows the fitting with the Fano line shape. The dotted lines for higher BE features are guide to eye. L shows the appearance of a new feature in the CIS spectra of FeAl and its position is marked by ticks in (b). To generate the feature L in the CIS plot of FeAl two different Fano line shapes are considered (denoted as Fano-1 and Fano-2) and are shown separately.}
%\label{fig4}
%\vskip -0.2cm
%\end{figure} 
 
The character of the VB features can be identified from the constant initial state (CIS) spectra as shown in figure 4(b). CIS spectra have been determined from figure 4(a) by plotting the intensity at the fixed BE positions in the VB as a function of photon excitation energy. The CIS plot is a result of interference of two simultaneously occurring processes: 1) the direct photoemission process which is weakly dependent on the incident photon energy and 2) the resonant process where there are excitations across the $3p-3d$ (in Fe) and $2p-3s$ (in Al) transitions. CIS spectra can be fitted (shown by solid lines in figure 4(b)) with a Fano line profile of the form $\sigma(h\nu)=\sigma_a \frac{(q+\epsilon)^2}{1+\epsilon^2} + \sigma_b$ and $\epsilon=(h\nu-E_0)/\Gamma$ \cite{Fano} where, $E_0$ is the resonance energy, $\Gamma$ is the half-width of the line. $q$ is the line profile index which represents the discrete/continuum mixing strength. The cross sections $\sigma_a$ and $\sigma_b$ represent different non-resonant background cross-sections.  

We find that the Fe $3d$ related features at -0.3 and -0.8~eV BE in figure 4(b), show an antiresonance dip which is consistent with the resonance observed in Fe metal \cite{Chandesris83, Kato85}. Fitting of -0.3~eV BE spectrum gives the Fano parameters $E_0$=~52.3~eV, $q$=~-0.38 and $\Gamma$=~2.5. Small value of $q$ which is characterized by a distinct antiresonance indicates a dominant itinerant character near E$_F$. Interestingly, at -2.2~eV BE, we have observed a new feature in the CIS spectrum marked by  $L$ in the figure 4(b) which appears to be overlying on the antiresonance dip and the intensity of this feature increases with the increasing BE from -2.2~eV to -4~eV. To fit the CIS spectra of -2.2~eV, -3~eV and -4~eV BE, two different Fano line shapes have been used such that the total CIS intensity=~$\alpha$~Fano-1~+~$\beta$~Fano-2 where $\alpha$ and $\beta$ are positive constants. The fitted parameters obtained for Fano-1 (dashed red line in figure 4(b)) and Fano-2 (dotted blue line in figure 4(b)) are quite different which gives the direct evidence of two different types of Fe $3d$ states in this system. We find that $q$ for Fano-1 is negative and is close to zero giving an antiresonance feature at 52 eV, whereas for Fano-2 it is quite large ranging from 10 to 2 between -2.2~eV and -4~eV BE spectra giving a resonance feature at 52 eV excitation. $E_0$ for Fano-1 ranges from 51.5 to 52.5~eV and for Fano-2 it ranges from 53.6 to 53.1~eV whereas $\Gamma$ for Fano-1 ranges from 2.2 to 3~eV and for Fano-2 it ranges from 0.4 to 1~eV between -2.2~eV and -4~eV BE spectra . The large value of $q$ and very low value of $\Gamma$ corresponding to the Fano-2 ($L$ in figure 4(b)) clearly indicates that a different kind of Fe $3d$ state with relatively localized character is also present in the sample. The above conclusions related to the different types of 3d states, are drawn based on results available in literature, where it is shown that for $q\longrightarrow0$, the Fano line shape shows a symmetric antiresonance behavior and this is primarily due to the transition from bound states to itinerant states\cite{Davis1, Davis2, Matshushita}. For the case $q>1$, one gets a resonance peak corresponding to transitions from bound states to relatively less itinerant states in the system\cite{Davis1, Davis2, Matshushita}.

In addition to the Fe $3p-3d$ transition, we have also observed a resonance feature at 73~eV excitation energy related to Al $2p-3s$ transition in FeAl (figure 4(b)), which is present all through the VB. This signifies that both Fe $3d$ and Al $3s-3p$ states are hybridized and present all over the VB of FeAl. The Fano parameters $E_0$, $q$ and $\Gamma$ for Al $2p-3s$ transition has been found to be in the range between 73 to 71.5~eV, -7 to -8 and 3.4 to 4~eV, respectively between -0.3~eV and -3~eV BE spectra. Similar large negative value of $q$ has also been observed for Al $2p-3s$ transition in Al-GaAs (110) surface \cite{Kobyashi}. 

The origin of two different values of $q$ in the CIS plots for the Fe states are attributed to the presence of Fe atoms in the system located at two different crystallographic sites and hence different environments. These two environments are as follows: a) the Fe in ordered FeAl system with eight nearest neighbor Al and 6 next nearest neighbor Fe atoms and b) Fe in a disordered neighborhood of Fe and Al atoms with eight near neighbors. The two types of environment of Fe atoms should have different contributions to the VB DOS. In a study on Mn$_2$As, Matshushita {\it et al.}, has reported that the two Mn atoms in the unit cell, which have different environments, are known to contribute in the VB DOS at different energy position and have different resonance character \cite{Matshushita}. The difference between our case and Mn$_2$As is that in Mn$_2$As, the two types of Mn atoms are located inside the same unit cell  with different environments, whereas in our case, the two types of Fe atoms are located in different phases of FeAl, one in an ordered B2 phase and the other in a disordered A2 phase.

%\begin{figure}
%\center
%\epsfxsize=100mm
%\epsfxsize=65mm
%\epsffile{Fig5.eps}
%\vskip -0.3cm
%\caption {Spin-unresolved and spin-resolved total [(a) and (e)] and partial DOSs of Fe $3d$ [(b) and (f)], Al $3p$ [(c) and (g)] and Al $3s$ [(d) and (h)] for both ordered and disordered phases, obtained by employing DFT within GGA exchange-correlation functional. Energy of states is given with respect to the Fermi level of the system.}
%\label{fig4}
%\vskip -0.2cm
%\end{figure}

In order to understand and support our experimental results, we have performed DFT based  electronic structure calculations for FeAl system in both ordered (B2) and structurally disordered phases. The  results of total (figure 5(a)) and partial DOS of Fe $3d$ (figure 5(b)), Al $3p$ (figure 5(c)) and Al $3s$ (figure 5(d)) for both ordered and disordered phases, obtained by employing DFT  within GGA exchange-correlation functional are shown and the respective spin-up and spin-down DOSs are shown in figures 5(e)-(h). It is quite evident from the results of partial DOS that the Fe $3d$, Al $3s$ and Al $3p$ states are strongly hybridized in the VB regions for both the ordered and disordered phases.

In case of ordered B2 phase, we obtain a ferromagnetic ground state with magnetic moment of $\sim$0.7~$\mu_B$ per formula unit (f.u.) which matches well with the theoretical calculations available in the literature \cite{Bogner, Williams, Chacham, Sundararajan, Kulikov}. The VB spectra of this system are mainly dominated by Fe $3d$ states which has major peak at around ~-1~eV.  The major contribution of Al $3p$ states to the valence spectra lies  in  the range from ~-2~ to ~-3.5~ eV (See figure 5(c)) whereas the Al $3s$ states contribute more to the DOS below ~-4.0~eV.  It has been observed in experiments that the ordered FeAl phase shows a paramagnetic ground state \cite{Wertheim,Huffman} with an effective moment  of ~1.5~ $\mu_B$.  To generate a nonmagnetic ground state in this system Mohn {\it et al.}, \cite {Mohn} used the LDA ~+~U approach with U =~5~eV (where $U$ is on-site Coulomb interaction).  In order to include the effect of strong correlation, if any, present in this system, we have also carried out the electronic structure calculations by using GGA~+~U method with U=~1-5~eV for $3d$ orbitals of Fe atom. With the introduction of non-zero values of U, the positions of the peaks of the Fe $3d$ states in the up spin channel shifted towards the Fermi level and their contributions to the DOS become equal to those of minority carriers for U=~5~eV. Thus, the system goes from FM to NM ground state due to introduction of the on-site Coulomb interaction. This change in the nature of ground state is similar to the one observed by Mohn et al.,\cite{Mohn}. However, it is to be noted that our experimentally observed paramagnetic phase could not be explained by the results obtained by both GGA and GGA~+~U calculations.

To probe the possible influence of disorder on the properties of FeAl system, we have also performed the calculations for the structurally disordered FeAl. For this purpose, we have constructed a (3$\times$3$\times$3) super cell of body centred cubic unit cell (like bulk Fe metal) in which Fe and Al atoms are randomly arranged. In this case, the ground state is again found to be ferromagnetic but with larger magnetic moment of ~1.89~$\mu_B$ per f.u. The results of total and partial DOS of disordered phase show that there is a shift in energies of Fe $3d$ and Al $3p$ peaks towards the lower binding energies as compared to the ordered phase. It is also observed that the value of DOS at the Fermi level in the disordered phase is less than that of the ordered FeAl phase. However, the total DOS (figure 5(a)) at $\sim$-1.3~eV and $\sim$-2.2~eV for the disordered phase is higher than those of the ordered FeAl phase.  We wish to mention here that, as opposed to the ordered B2 phase where the Fe and Al atoms are arranged in alternate positions in the lattice, there is a finite probability of having linear chains or formation of clusters made up of Fe atoms leading to a disordered phase. Thus, there will be much stronger exchange coupling between the Fe atoms in the disordered phase as compared to that of the ordered phase which in turn leads to a higher magnetic moment as is observed in the former.

%\begin{figure}
%\center
%\epsfxsize=100mm
%\epsfxsize=65mm
%\epsffile{Fig6.eps}
%\vskip -0.3cm
%\caption {The experimental difference spectra (which is shown as shaded region in green color) and the theoretical  PDOS of FeAl in both ordered (closed circles in red color) and disordered ( open circles in blue color) phases. Theoretical PDOS spectra are overlaid on the experimental difference spectra in order to identify the important features. Energy of PDOS is given with respect to the Fermi level.}
%\label{fig4}
%\vskip -0.2cm
%\end{figure}

To compare the theoretical and experimental results, the Fe $3d$ PDOS has been broadened and is shown in figure 6. A standard procedure adopted to broaden the DOS is mentioned in our earlier references \cite{Soma1}. To derive the information about the experimental Fe $3d$ PDOS the difference spectrum have been calculated by subtracting the off-resonance spectrum from the on-resonance spectrum. Similar procedure has been adopted by us to extract the information about Ce $4f$ PDOS in CeAg$_2$Ge$_2$ \cite{Soma1}, Co $3d$ and Mn $3d$ PDOS in Co$_2$MnSn \cite{Madhusmita}. To compare the experimental results with those obtained from the theoreictical calcualations, we have plotted the  experimental difference spectra (which is shown as green shaded region) and the theoretical  PDOS of FeAl in both ordered and disordered phases in figure 6.  In this figure, we have marked three features in the experimental difference spectra namely X, Y, and Z. We find the features X and Y resemble the Fe 3d peaks observed in ordered B2 FeAl phase.  On the other hand, the enhanced broad feature Z as observed in the experimental difference spectra looks distinctly different when it is compared with the theoretical PDOS  of ordered FeAl phase. In addition, we find that the feature Z match well with the theoretical PDOS of Fe 3d states in disordered FeAl phase. Thus, it is neceassary to include both the ordered and disordered phases in order to explain the observed experimental difference spectra. This observation also corroborates with  our XRD  results where we find the presence of finite disorder in this system. Furthermore, we also observe the presence of disordered FeAl in the system,  as discussed earlier, leads to the appearance of the extra feature (L)  in the CIS spectra. The shape of the resonance in the CIS spectra (figure 4(b)) shows the localized nature of feature L, which is also confirmed from the higher value of the magnetic moment (1.89~$\mu_B$) obtained from the theoretical calculations for the disordered system. The major peaks in DOS of disordered FeAl (figures 5 (a) and (b)) are found to lie much below the $E_F$ as compared to those in DOS of ordered FeAl which gives rise to the dominant itinerant character to the Fe 3d electrons in the former.

%\begin{figure}
%\center
%\epsfxsize=100mm
%\epsfxsize=65mm
%\epsffile{Fig7.eps}
%\vskip -0.3cm
%\caption {(a) Fe $3s$ core-level spectra of Fe metal and FeAl showing the experimental raw data as dots. Solid red and blue line show the total and the individual fit of core level peaks with the Tougaard background shown by green line. (b) XANES spectra of Fe metal and FeAl are shown on the top panel. In the bottom panel, the first derivative of the XANES spectra for Fe metal and FeAl are shown.}
%\label{fig4}
%\vskip -0.2cm
%\end{figure}

The evidence of the localized type moment in the system can be confirmed from the splitting of the Fe $3s$ core level. For Fe metal, it is reported that the $3s$ splitting remain identical in both the ferromagnetic as well as in the paramagnetic phase \cite{Fadley, Pickel}. Hence, the signature of the exchange splitting will give the information of the local moment present in the system \cite{Fadley, Pickel}. In figure 7(a) the exchange splitting ($\Delta$E) of Fe $3s$ peak for Fe metal and FeAl are shown. From the fitting $\Delta$E has been determined to be 4.3~eV and 4.0~eV for Fe metal and FeAl, respectively. The value of local magnetic moment can be calculated from the intensity ratio of the exchange split peak to the main peak.  For Fe metal and FeAl the intensity ratios are 0.4 and 0.32, respectively. This intensity ratio is proportional to the total spin by the relation S/(S+1) \cite{Qiu92} Hence, the magnetic moment per Fe atoms can be calculated using the relation $g\sqrt{S(S+1)}\mu_B$ where $g$ is the Lande $g$-factor. This gives the value of magnetic moment in Fe metal to be 2.11~$\mu_B$ and for FeAl to be 1.66~$\mu_B$. We find that the local moment determined from the exchange splitting of the Fe $3s$ core level (1.66~$\mu_B$) is in very good agreement with the effective moment obtained from the bulk AC-susceptibility measurement(1.5~$\mu_B$). Our calculated magnetic moment in ordered FeAl is 0.7~$\mu_B$ and in disordered FeAl is 1.89~$\mu_B$. Hence, with the coexistence of both these phases in the system, the observed moment is expected to lie in between these two magnetic moment values, which is observed in our case. Oku {\it et al.},\cite{oku06} obtained similar magnetic moment of $\sim$1.7$\mu_B$ in FeAl. They have attributed it to the moment present in the surface slab layers. However this does not seem to be the case, as is clearly concluded from our electronic structure and magnetic studies.

The unoccupied electronic states in FeAl have been studied by the XANES measurement at the Fe K-edge. The top panel in Fig. 7(b) shows the XANES spectra of Fe metal and FeAl. To extract the information about the unoccupied DOS, in the bottom panel of Fig. 7(b) we have shown the first derivative of the XANES spectra. We have observed two important differences between the XANES spectra of Fe and FeAl namely 1) shift in the main peak (from $\sim$7131~eV for Fe metal to $\sim$7140.5~eV for FeAl) corresponding to the transition from $1s$ to $4p$ levels (shown by arrows in Fig. 7(b)) and 2) increased absorption in the pre-edge region in the energy range between 7100~eV to 7125~eV in FeAl than in Fe metal. The shift of the $1s$ to $4p$ transition to higher energies observed in Fe molecules isolated in solid Ne \cite{Galoisy01} as compared to the pure Fe metal has been attributed to the presence of $3d^64s^2$ state rather than $3d^74s$ state in Fe metal \cite{Purdum82}. The presence of a higher vacancy in the Fe 3d state in FeAl is expected to increase the intensity of $1s$ to $3d$ absorption, which is observed in experiments. Comparing the Fe and FeAl pre-edge features we find that the position of the feature $M$ at $\approx$ 7112~eV shows no shift in energy indicating that the Fe states in FeAl is similar to Fe metal \cite{Galoisy01}. However, the feature $M$ in FeAl is narrow and enhanced in intensity than the Fe metal shows that the Fe $3d$ states are much more localized in this system. Feature $N$ in FeAl shows less intensity and is shifted towards lower binding energy, which gives the evidence that the Fe $3d$ states are strongly hybridized with the Al $3s-3p$ states which corroborates with our XPS and RPES results. Thus, the observations in XANES and magnetic measurement support the presence of localized and itinerant character of Fe in our sample as seen by PES.

\section{Conclusions}

The VB electronic structure of FeAl has been studied using RPES to understand the nature of Fe $3d$ states which is responsible for the magnetic behavior of this system. In FeAl, we find that two different kind of Fe $3d$ states are present, one with itinerant character and the other with less itinerant (localized) character. The itinerant character of the Fe $3d$ states in this system arises mainly due to its hybridization with the Al $3s-3p$ states and the evidence has been observed in both the core level and the XANES spectra. Magnetization and AC-susceptibility indicate the presence of magnetic clusters which could arise due to the finite disorder present in this system. First principles density of states calculations confirm the origin of the itinerant states due to the ordered $B2$ phase of FeAl and the origin of the less itinerant (localized) Fe $3d$ states due to the presence of the disordered phase of FeAl. The higher value of magnetic moment obtained in the calculation for the disordered phase (1.89~$\mu_B$) as compared to the ordered phase (0.7~$\mu_B$) also confirm its localized character. Experimentally obtained magnetic moment 1.6~$\mu_B$ also confirms the presence of disorder.

\section{Acknowledgement}
 
Dr. S. B. Roy is thanked for useful discussions on the magnetization data. Dr. S. K. Rai, Mr. A. Wadikar, Ms. Babita and Mr. V. K. Ahire are thanked for the experimental support. CK and AC thank Scientific Computing Group, RRCAT for their support. Glass blowing facility, RRCAT is thanked for the ampule preparation and vacuum sealing of the sample. DM thanks HBNI for financial support. The authors wish to thank Dr. P. D. Gupta, Dr. G. S. Lodha, Dr. S. K. Deb and Dr. P. A. Naik for their constant encouragement, support and discussion.\\

%\noindent $^*$Corresponding author. Tel.: +91-731-244-2596; Fax: +91-731-244-2140.\\
 % E-mail address: soma@rrcat.gov.in.\\
%\vskip -0.2cm
%\newpage
\vskip -0.6cm
%\bibliographystyle{elsarticle-num}  
%\begin{thebibliography}{00}

\newpage
\noindent {\large Figure Captions :}\\

\noindent Figure 1. (a) X-ray diffraction pattern of FeAl at 300 K recorded with synchrotron source at 19.3~keV energy. The experimental data are denoted by open red circles, while the black solid line through the circles represents the calculated pattern. The lower dotted line represents the difference curve between experimental and calculated patterns. The dots in (a) represent the small impurity phases present in the sample. Magnetization as a function of applied magnetic field at 300~K and 2~K is shown in (b). Zoomed region of M(H) curve as in (b) near H=~0 is shown in inset for the hysteresis loop observed at 2~K. AC-susceptibility as a function of temperature is shown in (c) with the Blocking temperature and SG freezing temperature marked as $T_B$ and $T_F$ respectively.\\

\noindent Figure 2. (a)Field Cooled (FC) and Zero Field Cooled (ZFC) magnetization measurement of FeAl under applied magnetic field of 0.01~T, 0.1~T and 1~T. (b) Temperature dependence of 1/$(\chi-\chi_0)$ for FeAl determined from the FC magnetization curves of 1~T and 0.1~T as shown in (a) (details are given in the text).\\

\noindent Figure 3. Comparison of the core-level spectra of FeAl with Fe metal and Al metal showing (a) Fe $2p$, (b) Fe $3p$, (c) Al $2s$ and (d) Al $2p$ core levels. Dotted lines show the positions of the core level peaks.\\

\noindent Figure 4. VB spectra of FeAl recorded across the Fe $3p-3d$ transition and the Al $2p-3s$ transition are shown in (a) with the prominent features marked as A, B, C. The constant initial state spectra of the VB features at fixed binding energies are shown as a function of photon energy in (b). The solid line shows the fitting with the Fano line shape. The dotted lines for higher BE features are guide to eye. L shows the appearance of a new feature in the CIS spectra of FeAl and its position is marked by ticks in (b). To generate the feature L in the CIS plot of FeAl two different Fano line shapes are considered (denoted as Fano-1 and Fano-2) and are shown separately.\\

\noindent Figure 5. Spin-unresolved and spin-resolved total [(a) and (e)] and partial DOSs of Fe $3d$ [(b) and (f)], Al $3p$ [(c) and (g)] and Al $3s$ [(d) and (h)] for both ordered and disordered phases, obtained by employing DFT within GGA exchange-correlation functional. Energy of states is given with respect to the Fermi level of the system.\\

\noindent Figure 6. The experimental difference spectra (which is shown as shaded region in green color) and the theoretical  PDOS of FeAl in both ordered (closed circles in red color) and disordered ( open circles in blue color) phases. Theoretical PDOS spectra are overlaid on the experimental difference spectra in order to identify the important features. Energy of PDOS is given with respect to the Fermi level.\\

\noindent Figure 7. (a) Fe $3s$ core-level spectra of Fe metal and FeAl showing the experimental raw data as dots. Solid red and blue line show the total and the individual fit of core level peaks with the Tougaard background shown by green line. (b) XANES spectra of Fe metal and FeAl are shown on the top panel. In the bottom panel, the first derivative of the XANES spectra for Fe metal and FeAl are shown.\\

\end{document}